\documentclass[10pt,a4paper]{article}

\usepackage[T1]{fontenc}
\usepackage[utf8]{inputenc}
\usepackage{amssymb,amsmath,amsthm,amsfonts}    
\usepackage{bm}                                 
\usepackage{mathtools}                          
\usepackage{stmaryrd}                           
\usepackage{breqn}                              
\usepackage[textfont=footnotesize,labelfont={footnotesize,bf}]{caption}
\usepackage{subcaption}
\usepackage{booktabs}
\usepackage[inline]{enumitem}
\usepackage{graphicx}
\usepackage{graphbox}
\usepackage[dvipsnames]{xcolor}
\usepackage[top=30mm,right=30mm,left=30mm,bottom=30mm]{geometry}
\usepackage[colorlinks=true,linkcolor=blue,urlcolor=blue]{hyperref}
\usepackage[affil-it,auth-sc]{authblk}
\usepackage[colorinlistoftodos,textwidth=25mm,textsize=footnotesize]{todonotes}
\usepackage{lipsum}                             
\usepackage{enumitem} 
\usepackage[section]{placeins}
\usepackage{tikz}
\usetikzlibrary{positioning}
\usepackage{scalefnt}
\usepackage[
	backend=biber,
	style=numeric,
	citestyle=numeric-comp,
	maxbibnames=99,
	minbibnames=99,
	maxcitenames=2,
	mincitenames=1,
	giveninits=true,
	sorting=none,
	sortlocale=auto,
	natbib=true,
	url=false,
	isbn=false,
	doi=true,
	eprint=true
]{biblatex}
\DeclareNameAlias{default}{family-given}
\DeclareUnicodeCharacter{02B9}{'}
\DeclareUnicodeCharacter{0308}{"}

\title{Flutter instability in solids and structures, with a view on biomechanics and metamaterials}

\author[1]{Davide Bigoni\footnote{Corresponding author: e-mail: \href{mailto:name@unitn.it}{bigoni@unitn.it}; phone: +39\,0461\,282507.}} 
\author[1]{Francesco Dal Corso} 
\author[2]{Oleg N. Kirillov}
\author[1]{Diego Misseroni}
\author[3]{Giovanni Noselli}
\author[1]{Andrea Piccolroaz}

\affil[1]{DICAM, University of Trento, via Mesiano 77, I-38123 Trento, Italy}
\affil[2]{Northumbria University, Newcastle upon Tyne NE1 8ST, United Kingdom}
\affil[3]{SISSA-International School for Advanced Studies, 34136 Trieste, Italy}

\date{}

\begin{document}

\maketitle

\begin{abstract}
The phenomenon of oscillatory instability called \lq flutter' was observed in aeroelasticity and rotor dynamics about a century ago. 
Driven by a series of applications involving nonconservative elasticity theory at different physical scales,  ranging from nanomechanics to the mechanics of large space structures and including biomechanical problems of motility and growth, research on flutter is experiencing a new renaissance.
A review is presented of the most notable applications and recent advances in fundamentals, both theoretical and experimental aspects, of flutter instability and Hopf bifurcation. Open problems, research gaps, and new perspectives for investigations are indicated. 
\end{abstract}

\paragraph{Keywords}
Hopf bifurcation \textperiodcentered\ 
Elasticity \textperiodcentered\ 
Non-holonomic constraints \textperiodcentered\ 
Non-Hermitian mechanics \textperiodcentered\ 
Non-conservative systems

\section{Introduction}

Flutter instability, a peculiar dynamic behaviour consisting of a blowing-up oscillatory motion, is a phenomenon which was discovered in structural mechanics almost a century ago as a consequence of the application of non-conservative, follower forces, to structures. The many strongly counter-intuitive features, characteristic of this instability, were highlighted by famous scientists, for instance, Vladimir Bolotin and Warner Koiter, and \textcolor{black}{inspired} a strong research effort on this topic. 
The purpose of this review is to show how, although initially confined to slender rods, the heart of structural mechanics,  the research on flutter instability has been spread through the years along various directions and across a broad range of physical scales. In the following sections, it is shown how this intriguing mechanical instability is displayed in several different scientific and technological fields. 
Indeed, in mechanobiology follower forces were found to act on the human spine, and flutter instability was shown to explain circumnutations during the growing of plant shoots, motility of cells through eukaryotic cilia, and vertebral segmentation in embryos. Moreover, flutter instability affects control problems, the design of soft robotic actuators, graphene peeling, wire drawing, plastic extrusion, sheet paper printing, hoist and elevator systems, and deployment and retrieval of space tether systems, and solar sails. 
These topics are briefly reviewed below, while the famous problem of aeroelastic flutter, which leads to the structural failure of flying vehicles (but also chimneys and bridges) and, accordingly, affects strongly their design requirements \citep{ ED2021, hodges, MM2010, MP2016, TV2019}, is not included \textcolor{black}{as it would require a separate treatment. Nevertheless,  similarities between flutter in structural systems and in fluid-structure interaction problems are highlighted}. 

In a detailed vein, new achievements are presented in the {\it mechanics} of flutter, including the experimental approach to follower forces, internal activity, effects related to non-smoothness of the equations of motion, non-holonomic constraint, and, finally, the extension from structural flutter to flutter of a continuum, as related to non-Hermitian materials.
Before proceeding to review flutter in all of these fields, it is instrumental to provide a short presentation of flutter instability and its basic features.

\subsection{The counter-intuitive character of flutter instability}

\begin{figure}[!ht]
\centering\includegraphics[width=\textwidth]{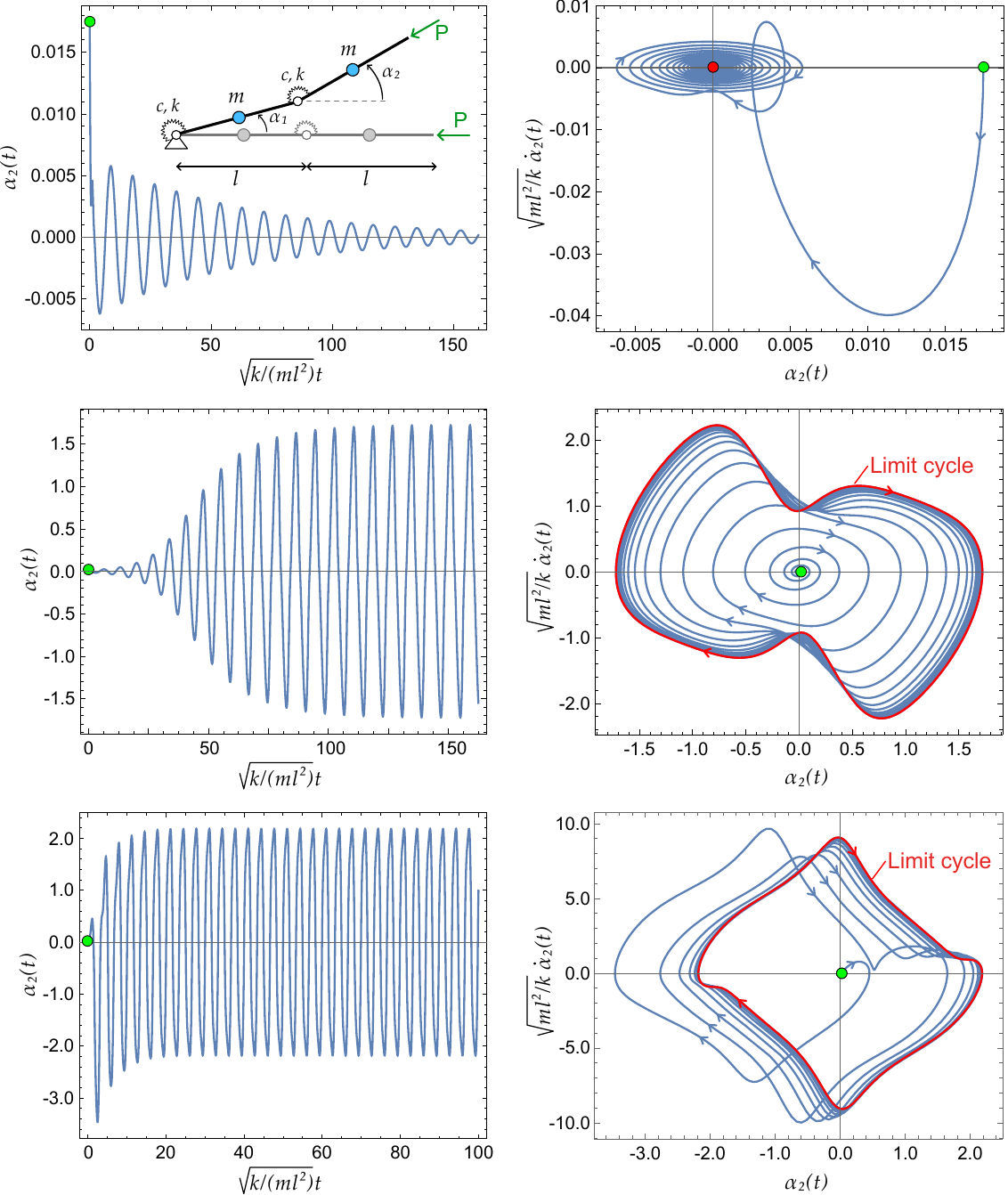}
\caption{
Results reported in the figure refer to the \lq Ziegler's double pendulum' (shown in the inset, upper part on the left), a 2 d.o.f. ($\alpha_1$, $\alpha_2$) two-bar structure with two concentrated masses $m$ and two linear viscoelastic hinges with elastic stiffness  $k$ and viscous damping $c$.
Evolution of the coordinate $\alpha_2(t)$ (left parts) and its phase-portrait (right parts) for the \lq Ziegler's double pendulum' with $c=0.1\sqrt{k m l^2} $, obtained from the numerical integration of the nonlinear equations of motion (\ref{nonlinearazzostocazzo}), with the initial conditions $\alpha_1(t=0)=\dot\alpha_1(t=0)=\dot\alpha_2(t=0)=0$ and $\alpha_2(t=0)=1^\circ$, identified as a green spot. 
Results reported at different horizontal lines differ in the follower load value $P=\{0.5,1.5,4\}P_f(c)$, increasing from the upper to the lower parts. The final state ($t\rightarrow\infty$) corresponds to the rest condition (red spot) for $P=0.5P_f(c)$ (upper line) and to the achievement of an oscillatory limit circle (red closed curve) in the other two cases, $P=1.5P_f(c)$ and $P=4P_f(c)$ (centre and lower line).
}
\label{flutter1}
\end{figure}

Flutter is a dynamic instability that manifests itself as an oscillatory motion of blowing-up amplitude which, in certain cases, may be interpretable as a Hopf stable bifurcation \citep{NQS2000} when eventually reaches a limit cycle. 
The simplest structural system displaying flutter instability is the \lq Ziegler's double pendulum' \citep{BL2018,Z1977,Z1952} (inset of Fig.~\ref{flutter1}, upper part on the left), \textcolor{black}{which is a 2 d.o.f. system and} mainly differs from the classical double pendulum in the external loading condition. In more detail, \textcolor{black}{the two degrees of freedom $\alpha_1$ and $\alpha_2$ correspond to the rotation of} two rigid bars (each of length $l$ and containing a concentrated mass $m$ at its centre point, which is a particular case of a more general and original Ziegler's model \cite{Z1952}), connected together and to a fixed point through linear viscoelastic hinges, having elastic stiffness $k>0$ and viscous damping  $c\geq 0$. \textcolor{black}{The external load is applied to}  the free end of the second rigid bar\textcolor{black}{, however, differently from the classical double pendulum, the load $P$} has the peculiarity to remain aligned parallel to the tangent to the structure at its application point, because assumed to be {\it tangentially follower}.
This type of loading has  {\it non-conservative} nature, so it can produce a non-null work along a closed-loop of deformation. \textcolor{black}{The occurrence  of a dynamic instability, flutter, in the absence of quasi-static bifurcation, is now shown for this simple non-conservative system.}

\textcolor{black}{From the principle of virtual power,} the nonlinear equations governing the motion of the  \lq Ziegler's double pendulum' can be derived as
\begin{equation}
\begin{array}{ll}
\label{nonlinearazzostocazzo}
\displaystyle 
\frac{5}{2}\ddot{\alpha}_1 
+  \ddot{\alpha}_2 \cos{(\alpha_1-\alpha_2)} 
+  
\left(
\dot{\alpha}_2^2 
-\frac{2P}{ml} \right)
\sin{(\alpha_1-\alpha_2)} 
+ \frac{2c}{ml^2} (2\dot{\alpha}_1-\dot{\alpha}_2)
+ \frac{2k}{ml^2} (2\alpha_1-\alpha_2)  = 0, \\ [5 mm]
\displaystyle 
\ddot{\alpha}_1 \cos{(\alpha_1-\alpha_2)} + \frac{1}{2} \ddot{\alpha}_2  
-\dot{\alpha}_1^2 \sin{(\alpha_1-\alpha_2)} 
- \frac{2c}{ml^2} (\dot{\alpha}_1-\dot{\alpha}_2) 
 - \frac{2k}{ml^2} (\alpha_1-\alpha_2) = 0 ,
\end{array}
\end{equation}
\textcolor{black}{and their linearization  about the straight configuration $\alpha_1=\alpha_2=0$ is expressed by
\begin{equation}\label{zieglinear}
\boldsymbol{M}\ddot{\boldsymbol{q}}
+
\boldsymbol{D}\dot{\boldsymbol{q}}
+
\left(\boldsymbol{K}+P\boldsymbol{\mathsf{K}}^{(P)}+P\boldsymbol{\mathsf{N}}^{(P)}\right)\boldsymbol{q}=\boldsymbol{0},
\end{equation}
where $\boldsymbol{q}= \left\{\alpha_1,\alpha_2\right\}^T$ is the vector collecting the Lagrangian coordinates, $\boldsymbol{M}$, $\boldsymbol{D}$, and $\boldsymbol{K}$ are respectively the  symmetric mass, damping, and stiffness matrices of the pendulum,
\begin{equation}
\boldsymbol{M}=
\left[
\begin{array}{ccc}
5 & 2\\
2 & 1   
\end{array}
\right
]\dfrac{m l^2}{4},
\qquad
\boldsymbol{D}=
\left[
\begin{array}{ccc}
2 & -1\\
-1 & 1    
\end{array}
\right
]c,
\qquad
\boldsymbol{K}=
\left[
\begin{array}{ccc}
2 & -1\\
-1 & 1   
\end{array}
\right
]k,
\end{equation}
while $P\boldsymbol{\mathsf{K}}^{(P)}$ is an additional symmetric stiffness  matrix and $P\boldsymbol{\mathsf{N}}^{(P)}$ a skew-symmetric circulatory matrix, the sum of which introduces the effect of the follower  force $P$, 
\begin{equation}
\boldsymbol{\mathsf{K}}^{(P)}=
-\left[
\begin{array}{ccc}
2 & -1\\
-1 & 0   
\end{array}
\right
]\dfrac{l}{2},
\qquad
\boldsymbol{\mathsf{N}}^{(P)}=
\left[
\begin{array}{ccc}
0 & 1\\
-1 & 0    
\end{array}
\right
]\dfrac{l}{2}.
\end{equation}
}
\textcolor{black}{In terminology of the work \cite{H2015}, the \lq Ziegler's double pendulum' is a MDKN-system due to absence of gyroscopic forces, which are typical in the reduced order models of fluid-structure interaction problems of aeroelasticity \cite{MP2016, ED2021, MM2010} thus representing MDGKN systems \cite{H2015}.}
\textcolor{black}{Interestingly, the equilibrium of the structure is only possible in the straight state, $\boldsymbol{q}=0$, where the viscoelastic hinges are undeformed. 
Indeed,  any quasi-static bifurcation can be excluded as
\begin{equation}
    \mbox{det}\left[\boldsymbol{K}+P\boldsymbol{\mathsf{K}}^{(P)}+P\boldsymbol{\mathsf{N}}^{(P)}\right]=k^2\neq 0,\qquad \,\,\forall{P}.
\end{equation}
Nevertheless, the investigation of the linearized dynamic response discloses the possibility for the trivial configuration to become unstable within a certain set of values of the follower load $P$.
More specifically, the} following observations can be drawn:
\begin{itemize}

    \item \textcolor{black}{Considering small amplitude harmonic vibrations in time $t$ around the straight configuration, $\boldsymbol{q}= \left\{a_1,a_2\right\}^T\,e^{-i\Omega t}$, with  amplitudes  $a_{1,2}$ and  angular frequency $\Omega$, the linearized equations of motion \eqref{zieglinear} lead to the following eigenvalue problem
    \begin{equation}
    \left[-\Omega^2\boldsymbol{M} - i \Omega \boldsymbol{D} + \left(\boldsymbol{K}+P\boldsymbol{\mathsf{K}}^{(P)}+P\boldsymbol{\mathsf{N}}^{(P)}\right)\right]\left\{a_1,a_2\right\}^T=\boldsymbol{0}.
    \end{equation}
    At varying values of the follower load $P$, the  eigenvalues $\Omega^2_{1,2}$ change their nature and therefore the small amplitude vibration behaves differently, in particular:
    \begin{enumerate}[label=(\roman*)]
        \item when the eigenvalues $\Omega_{1,2}^2$ are real and positive and $c>0$, the motion is characterized by an amplitude exponentially decaying towards the trivial configuration, which is therefore asymptotically stable. In the case when $c=0$, the motion is oscillatory with a constant amplitude around the trivial configuration, which is therefore  stable;
        \item when the eigenvalues $\Omega_{1,2}^2$ are complex conjugate, the motion is oscillatory around the trivial configuration with exponentially increasing amplitude. Therefore, the trivial configuration is unstable and \lq flutter' occurs;
        \item when the eigenvalues $\Omega_{1,2}^2$ are real and negative, the motion has an exponentially increasing amplitude. As in the previous case, the trivial configuration is unstable, but  \lq divergence' occurs in the absence of oscillations.
    \end{enumerate}
    }
    Under the assumption of null viscosity, $c=0$, the linearized analysis reveals that the three above regimes correspond  to 
    \begin{enumerate}[label=(\roman*)]
        \item $P<P_f^*$;
        \item $P_f^*<P<P_d^*$;
        \item $P>P_d^*$;
    \end{enumerate}
    where $P_f^*$ and $P_d^*$ are respectively the threshold (positive) values for achieving flutter and divergence in the non-dissipative  system ($c=0$), 
    \begin{equation}
    P_f^*=3\frac{k}{l},\qquad P_d^*=\frac{13}{3} \frac{k}{l}.
    \end{equation}
    However, when viscosity $c$ is kept into account, the trivial configuration is stable for $P<P_f(c)$, where $P_f(c)$ is the critical load for flutter  in the presence of non-null damping, 
    \begin{equation}
    \label{viscosazzo}
    P_f(c)=\left(\frac{39}{22}+\frac{4c^2}{3kml^2} \right)\frac{k}{l}.
    \end{equation}

    \item The comparison between the flutter threshold  $P_f^*$ for the non-dissipative system with the corresponding value  $P_f(c=0)$ for the dissipative system with vanishing damping constant shows the so-called {\it Ziegler-Bottema destabilization paradox}  \citep{Abdullatif2018, Bottema1956, HG2003, KDM2007, K2021, KV2022, KV2010, Kirillov2005, Kirillov2004,  Luongo2015b, Luongo2015, M2019, OReilly1996, Z1952}. The paradox consists in the surprising fact that the limit of null viscosity ($c \longrightarrow 0$) leads to a critical flutter load $P_f(c=0)=39k/(22l)$ smaller (much smaller, in the case of the example) than that obtained for the non-dissipative system, $P_f^*=3k/l$. This vivid example of a \lq dissipation-induced instability' \citep{KDM2007,K2021,KV2022,marsden} is strongly counter-intuitive. 

    \item Considering a non-null viscosity $c>0$, the nonlinear numerical solution of equations (\ref{nonlinearazzostocazzo}) evidences a decaying oscillatory motion in case (i), an oscillatory motion with an initially blowing-up amplitude eventually displaying a limit cycle in case (ii), and an oscillatory motion of large amplitude, again displaying a limit cycle in case (iii). These three types of dynamic response can be appreciated from   Fig. \ref{flutter1}, where the evolution in time of the coordinate $\alpha_2(t)$ and its phase-portrait are reported for the \lq Ziegler's double pendulum', with $c=0.1\sqrt{k m l^2} $ and under three different values of the follower load $P$, equal to $\{0.5,1.5,4\}P_f(c)$, so that all cases (i.)--(iii.) are explored. Note that the flutter motion reported in the centre line of Fig. \ref{flutter1} occurs for a value of load $P=1.5P_f(c)<P_f^*$ smaller than $P_f^*$, obtained from the analysis of the non-dissipative system.

\end{itemize}

\begin{figure}[!h]
\centering\includegraphics[width=\textwidth]{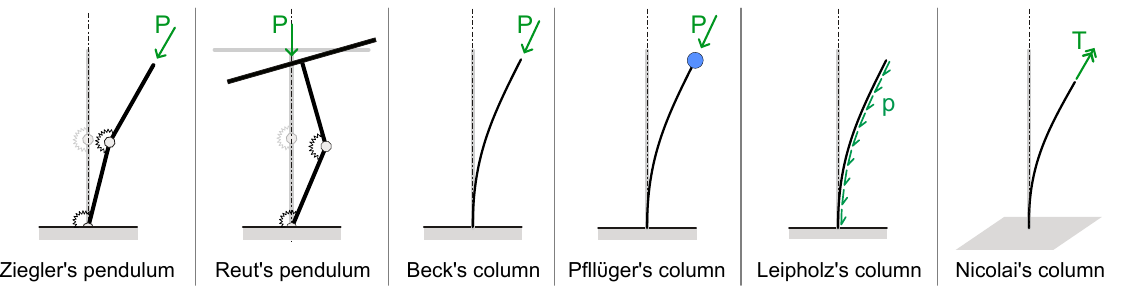}
\caption{Discrete and continuous structural models displaying flutter instability when subject to nonconservative loads. All loads are tangentially follower, except in the case of the Reut's pendulum.}
\label{cases}
\end{figure}

Further examples of simple nonconservative structural systems displaying flutter instability have been introduced in the first part of the last century. They are often included in scientific textbooks on structural stability \citep{bigonilibro,B1963,cedolin2010stability,Luongo2023}. Among the many, the following ones are recalled (Fig.~\ref{cases}):
\begin{itemize}
\item Reut's double pendulum \citep{R1939}, differing from the \lq Ziegler's double pendulum' only in the load application. More specifically, the constant magnitude load $P$ slides along a rigid bar to maintain its  application line always coincident with the undeformed state;
\item Beck's column \citep{B1952}, an elastic cantilever rod differing from the \lq Ziegler's double pendulum' only for the continuous and uniform distribution of mass, bending stiffness and dissipation;
\item Pfluger's column \citep{Pfluger1955,Pfluger1950}, which enhances Beck's column by considering the presence of a lumped mass at the application point of the follower load;
\item Leipholz's column \citep{L1987,LP1984}, an elastic cantilever rod subject to a uniform distribution of follower tangential load acting along its axis;
\item Nicolai's column \citep{nicolai1929,nicolai1928}, an elastic cantilever rod with a follower twist load applied at the free end, so that three-dimensional motion is realized, instead of a planar one.
\end{itemize}

Interestingly, these structural models do not only display flutter instability but, similarly to the \lq Ziegler's double pendulum',  also counterintuitive behaviors, often referred as \lq paradoxes'  \cite{Abdullatif2019,bolotinarticle,Jacoby1986,Langthjem2000,Kirillov2019,Lerbet2013,Lottati1987,LFD2016, KO2021, Seyranian2014, B2020rcd, B2020tam, B2021}.

Recent theoretical works include studies of the integrability of the \lq Ziegler's double pendulum' \citep{P2023} and in general of mechanical systems with non-potential positional (circulatory) forces \citep{K2022rcd,K2022umn}, destabilizing influence of circulatory forces on a degenerate equilibrium of a potential system \citep{B2021,B2020rcd,B2020tam}, delay of flutter onset \citep{NT2021} and parametric instabilities in the \lq Ziegler's double pendulum' with the periodic follower load \citep{BL2018}, and stability of systems with follower forces under kinematic constraints \citep{L2020}.

The concept of flutter instability, exemplified with the nonconservative structures reported in Fig. \ref{cases}, was easily extended to a range of physical scales much broader than the scale considered when the simple structures were invented already a century ago.
Indeed, flexible slender structures are ubiquitous both in engineering applications and in biological systems. Therefore, it is not surprising that studies on fluttering of solar sails \citep{D2011} parallel others on flutter of flagella induced by dyneins inside eukaryotic cells \citep{BD2016,DCLG2017}.

\subsection{\textcolor{black}{Similarities between flutter from follower loads and from fluid-structure interaction }}\label{aeroelasticity}
\textcolor{black}{Although flutter instabilities arising from fluid-structure interaction are not the focus of the present article, the common nature shared by the equations governing the motion of the \lq Ziegler's double pendulum' and of aeroelastic systems is now highlighted.\\
The equations of motion of a structural system in a fluid can be expressed by \cite{H2015,K2021}
\begin{equation}\label{aeroelasticitylinear}
\boldsymbol{M}\ddot{\boldsymbol{q}}
+
\boldsymbol{D}\dot{\boldsymbol{q}}
+
\boldsymbol{K}\boldsymbol{q}=\boldsymbol{F}_a ,
\end{equation}
where the vector $\boldsymbol{q}$ collects the Lagrangian coordinates, $\boldsymbol{M}$, $\boldsymbol{D}$, and $\boldsymbol{K}$ are respectively the symmetric mass, damping, and stiffness matrices of the structure under consideration, while  $\boldsymbol{F}_a$ is the aeroelastic force modelling the interaction between the structure and the fluid. The fluid is assumed in a laminar flow at constant velocity $U$, along a fixed direction, and in a quasi-steady state. The latter  assumption holds when the structural  vibrations 
can be neglected as 
characterized by a velocity much smaller than that of the fluid flow. 
Under this circumstance, the  coupling  between the structure and the fluid reduces to a weak coupling and so the aeroelastic force $\boldsymbol{F}_a$ can be approximated at small amplitude vibrations as \cite{Luongo2023,MP2016,ED2021}
\begin{equation}\label{forceexp}\boldsymbol{F}_a=\boldsymbol{F}_a^0-
U\left(\boldsymbol{\mathsf{D}}^{(U)}+\boldsymbol{\mathsf{G}}^{(U)}\right)\dot{\boldsymbol{q}}
-
U^2\left(\boldsymbol{\mathsf{K}}^{(U)}+\boldsymbol{\mathsf{N}}^{(U)}\right)\boldsymbol{q}. 
\end{equation}
The approximated expression \eqref{forceexp} for the aeroelastic force $\boldsymbol{F}_a$ can be obtained by considering that (i)  the aerodynamic coefficients vary in time due to change of the attack angle of the laminar flow with the structure and (ii) the kinetic pressure of the fluid on the structure is referred to the relative velocity of the former with respect to the latter. It follows that the aeroelastic force $\boldsymbol{F}_a$ is given by the summation of three contributions: a constant term   $\boldsymbol{F}_a^0$ and two terms linear in the position $\boldsymbol{q}$ and in the velocity $\boldsymbol{\dot q}$. The latter two contributions are specified through the symmetric damping matrix $U\boldsymbol{\mathsf{D}}^{(U)}$, the skew-symmetric gyroscopic matrix $U\boldsymbol{\mathsf{G}}^{(U)}$, the symmetric stiffness matrix $U^2\boldsymbol{\mathsf{K}}^{(U)}$, and the skew-symmetric circulatory matrix $U^2\boldsymbol{\mathsf{N}}^{(U)}$.
By neglecting the constant contribution $\boldsymbol{F}_a^0$ as responsible  only in  a shifting of the vibration centre, the equations of motion \eqref{aeroelasticitylinear} can be rewritten as an MDGKN-system \cite{H2015}
\begin{equation}\label{aerolinear}
\boldsymbol{M}\ddot{\boldsymbol{q}}
+
\left(\boldsymbol{D}+U\boldsymbol{\mathsf{D}}^{(U)}+U\boldsymbol{\mathsf{G}}^{(U)}\right)\dot{\boldsymbol{q}}
+
\left(\boldsymbol{K}+U^2\boldsymbol{\mathsf{K}}^{(U)}+U^2\boldsymbol{\mathsf{N}}^{(U)}\right)\boldsymbol{q}=\boldsymbol{0},
\end{equation}
disclosing that the effect of aerodynamic force can be interpreted as a change in the stiffness, similarly to the case of the \lq Ziegler's double pendulum', but also as a change in damping. It follows that dynamic instabilities similar to flutter of the \lq Ziegler's double pendulum' can be reached by tuning the driving parameter of the flow, corresponding to the velocity $U$, which replaces in a sense the follower force $P$. In general, the presence of gyroscopic forces in MDGKN-systems allows to treat them as Hamiltonian systems perturbed by dissipative and nonconservative positional forces with similar descriptions of the destabilizing effect of these forces via singularities on the stability boundary \cite{KHR2007,KDM2007,K2021} as was first done by Bottema \cite{Bottema1956} for the \lq Ziegler's double pendulum'.
} 

\section{Flutter in engineering systems}

\subsection{Graphene peeling}

\textit{Graphene peeling} is a phenomenon that occurs when a graphene sheet detaches from a substrate due to external forces or internal stresses which arise from the adhesion between the graphene sheet and the substrate. When the substrate is deformed or bent, the graphene sheet can follow the substrate motion and peel off due to the release of the adhesion forces \citep{B2020,BB2022}. One of the mechanisms that can trigger graphene peeling is flutter instability induced by follower forces, which are caused by the interaction of the graphene sheet with a fluid flow or a moving substrate. In that case, flutter instability can induce a series of deformations in the graphene sheet, including waves, ripples, and wrinkles, which can amplify stresses and promote the detachment of the sheet. Understanding the mechanisms of graphene peeling induced by flutter instability and follower forces is essential for the design and optimization of graphene-based devices and materials.

\subsection{Moving structures}

The recently published book by Banichuk et al. \citep{Betal2020} is devoted to the stability analysis of initially straight and axially moving structures \citep{SV2023} such as strings, beams, plates or membranes. The structures are studied in terms of modal analysis, transverse vibrations, or stability and buckling, with applications for instance to paper production. Owing to the small flexural stiffness and low specific mass, paper webs are particularly susceptible to self-excited vibrations caused by a fluid-structure interaction with the surrounding medium \citep{TV2019}. 
This is also true when a pinned-free beam embedded in an axial fluid flow, is subject to feedback-based actuation \citep{Aspelund}. 

An interesting case of friction-induced self-excited instability of an axially moving structure is investigated by Spelsberg-Korspeter et al. \citep{SKH2008}. The model comprises an Euler–Bernoulli beam that moves across a spatial domain with linear guides at the boundaries and is in point-contact with frictional pads in the middle; the setup serves as a prototype problem for break squeal. The beam thickness is consistently taken into account through careful modelling of the contact kinematics. Divergence and flutter instabilities are investigated by means of a stability analysis with perturbation techniques revealing the Ziegler-Bottema destabilization phenomenon at vanishing damping \citep{HG2003,K2021, KV2022,KV2010,LFD2016, M2019, Z1952}.

\subsection{Structures with time-varying length and energy harvesters}

Dynamic instabilities in axially moving structures of \textit{time-varying length} are encountered in numerous technical applications, such as wire drawing, plastic extrusion, industrial robots, sheet paper printing, hoist and elevator systems \citep{SV2023}, and deployment and retrieval of space tether systems \citep{BL1993}, including solar sails \citep{D2011, Li2017, Lu2021, Lu2022}. Elastic rods of varying lengths may be subject to configurational forces \citep{BigoniEsh}, which may concur with follower forces to create complex structural behaviours. Varying length elements could be used as deployable structures for energy harvesting, a field where the research on flutter instability is flourishing, due to the possibility of converting a phenomenon usually considered harmful into a profit \citep{allen2001energy, barrero2010energy, doare2019dissipation, doare2011piezoelectric, li2022energy, tang2009coupled}.

\subsection{Flutter in granular materials and solids}

Self-excited vibrations and dynamical instabilities have long been known to occur during sliding contact {\it with friction} between solids \citep{den, ibra1, ibra2}; examples being the violin string excited by a bow, the brake \lq squeal', the \lq chatter' produced by a cutting tool of a machine, the \lq song' of a fingertip moved upon the rim of a glass of water \cite{K2021}. 
The usual explanation for these friction-related instabilities is \lq stick and slip', an alternate switch between static and kinetic friction \citep{den1, den, riceruina}. 

Assuming constant (time-independent) Coulomb friction, flutter instability has been theoretically proven for an elastic half-space sliding with constant velocity against a rigid and frictional constraint \citep{adams, martinsb, nguyenb, simmar}. 
Sliding with friction is a problem of great relevance in several fields and flutter may have important consequences on unstable fault slip in the Earth's crust, which generates earthquakes. 

In parallel to these works, elastoplastic continua characterized by nonassociative flow rule, the counterpart of Coulomb friction for solids, have been demonstrated to display blowing-up unstable waves, as related to the frictional behaviour of the material \citep{bigoniloret, bigoniwillis, bigoniflutt, loretsimoes,loret92, piccolroaz,rice77}. 
The elastoplasticity constitutive equations  are written in rate form so that they relate the rate of the Cauchy stress $\dot{\boldsymbol{\sigma}}$ to the rate of strain $\dot{\boldsymbol{\varepsilon}}$ as
\begin{equation}
\label{platico}
\dot{\boldsymbol{\sigma}} = 
\mathcal{E} 
\dot{\boldsymbol{\varepsilon}}
- 
\frac{
<\dot{\boldsymbol{\varepsilon}}
\cdot 
\mathcal{E} 
\boldsymbol{Q}>}{H+
\boldsymbol{Q}\cdot\mathcal{E} 
\boldsymbol{P} 
}
\mathcal{E} 
\boldsymbol{P}, 
\end{equation}
where $\mathcal{E}$ is the fourth-order elastic tensor, $H$ the hardening modulus (null for perfect plasticity, and negative for softening), $\boldsymbol{Q}$ and$\boldsymbol{P}$ the yield function gradient and the flow-mode tensor, respectively. 
Note that the symbol $<\cdot>$ is the Macaulay bracket operator, defined for every scalar $\alpha$ as $<\alpha> = (\alpha - |\alpha|)/2$. 
This operator is crucial to understand that the rate equations (\ref{platico}) are piece-wise linear and differentiate the plastic loading response
\begin{equation}
\label{platico2}
\dot{\boldsymbol{\sigma}} = 
\mathcal{C} 
\dot{\boldsymbol{\varepsilon}}, 
~~~~
\mathcal{C} = 
\mathcal{E} 
- 
\frac{1}{H+
\boldsymbol{Q}\cdot\mathcal{E} 
\boldsymbol{P}}
\mathcal{E} 
\boldsymbol{P}
\otimes 
\mathcal{E} 
\boldsymbol{Q}
, 
\end{equation}
where $\mathcal{C}$ is the fourth-order elastoplastic \lq in-loading' tensor, occurring when $\dot{\boldsymbol{\varepsilon}} \cdot \mathcal{E} \boldsymbol{Q} > 0$, from elastic unloading (and neutral loading) response
\begin{equation}
\label{elatico}
\dot{\boldsymbol{\sigma}} = 
\mathcal{E} 
\dot{\boldsymbol{\varepsilon}}, 
\end{equation}
occurring when $\dot{\boldsymbol{\varepsilon}} \cdot \mathcal{E} \boldsymbol{Q} < 0$ (when $\dot{\boldsymbol{\varepsilon}} \cdot \mathcal{E} \boldsymbol{Q} = 0$). For nonassociative plasticity, where a scalar $\beta$ does not exist such that $\boldsymbol{P} = \beta \boldsymbol{Q}$, the elastoplastic operator lacks symmetry and a boundary value problem becomes nonself-adjoint. 
With reference to the elastoplastic \lq in-loading' tensor, equation (\ref{platico2}), the acoustic tensor $\boldsymbol{A}$, with components
\begin{equation}
\label{acuplat}
    A_{ij}(\boldsymbol{n}) = \mathcal{C}_{ihjk}n_hn_k,
\end{equation}
defines the propagation in the direction $\boldsymbol{n}$ of a plane (or an acceleration) wave characterized by plastic-loading. 

By finding this lack of symmetry for the acoustic tensor in the elastoplastic constitutive modelling of granular materials, Rice \citep{rice77} first understood that this asymmetry may lead to the counterpart of flutter instability in a continuum, occurring when the acoustic tensor possesses two complex conjugate eigenvalues. 

The blowing-up of a signal during its propagation in a solid may become possible in materials characterized by the fourth-order tensor (\ref{platico2}), because elastoplasticity with nonassociative flow rule has been shown to produce energy in closed loading cycles \citep{petryk}. This circumstance does not necessarily violate the conservation of energy, because a release of energy can be produced at the expense of the strain energy stored in the material in connection with the presence of initial prestress, always needed to generate plastic flow.

The analysis of constitutive equations for granular materials \citep{gajo} reveals that flutter instability is a common feature, a fact in agreement with the observation that granular matter is prone to unstable releases of energy. 
However, experimental results indisputably showing flutter in a continuum are still unavailable, but evidence has been found of oscillatory instabilities occurring in different situations involving the mechanics of granular materials and pointing to flutter instability.
Examples are the \lq singing sand'  \citep{andreotti, dag}, the sound emissions during straining of snow \citep{pati}, and the \lq silo music' \citep{muite}. 

The analogy between flutter in a continuum and in a structure is more strict than it may appear, in fact, in both cases: 
(i.) flutter initiates when a complex conjugate eigenvalue solution for vibrations emerges, which corresponds to an oscillation of increasing amplitude; 
(ii.) a necessary condition for flutter is the lack of symmetry (or self-adjointness) in the mechanical system; 
(iii.) in a space of parameters, flutter occurs usually within a region, which separates stability from divergence, the latter being an exponentially blowing-up motion; 
(iv.) flutter ultimately leads to a self-sustaining oscillation, which absorbs energy from a steady source. 
This analogy will be invoked later in Section  4\ref{apezzi} to shed light on the limitations inherent to linearized stability analyses based on the plastic branch of the operator, equation (\ref{platico2}).

\section{Flutter in biology at different scales}

\subsection{Motility and propulsion}

A characteristic example is provided by eukaryotic cilia or flagella, hair-like active structures that protrude from many types of cells and act as a fundamental unit of motion, converting chemical energy into mechanical work in the form of an oscillatory beating state, leading to the propulsion of motile cells \citep{GBP2020,LGK2018}. The beating dynamics of `ciliary carpets' is prone to synchronization, both facilitating locomotion and navigation of single-celled peritrichous microorganisms and pumping mucous and other circulating fluids in the brain and lungs of animals \citep{Setal2022,GBP2020}. Fluid-structure interaction defines the dynamics of such structures  \citep{Setal2022,W2022,TV2019}. 

Eukaryotic cilia and flagella are characterized by a highly conserved and intricate internal structure (axoneme) in which dynein molecules (molecular motors)  collectively exert forces on cytoskeletal microtubule doublets leading to periodic oscillations \citep{BD2016, nicastro_2018}. However, the mechanism that regulates this molecular machinery is not fully understood \citep{LGK2018,V2019nl}. Mathematical models have been proposed to explain the bending motion of flagella in terms of intrinsic feedback mechanisms that regulate the motor activity of dyneins \citep{bayly_2015, lesich_2010,  howard_2016}. Interestingly, Bayly and Dutcher proposed a mechanism that does not require feedback regulation of dynein activity to generate oscillatory motions, using an elaborate model of the axoneme \citep{BD2016}. The proposed mechanism relies on a dynamic flutter instability induced by the internal dynein forces, which turn out to be \textit{follower}, i.e.\ always aligned along the centerline of the deforming filament \citep{B1963,Z1952}. De Canio et al. \citep{DCLG2017} simplified such a model by considering a single microfilament, confined to planar motion in a viscous fluid while clamped at one end and acted on by a molecular motor exerting a point follower force at the free end. These authors were able to demonstrate a  bifurcation of the filament to beating via flutter. Remarkably, linearization in \citep{DCLG2017} reduces to the classical \textit{Beck column} \citep{B1952} with an external damping \citep{T2016}, where the acceleration term is neglected. Ling et al.\ \citep{LGK2018} extended the nonlinear model by including a \textit{distributed} follower load as in the classical \textit{Leipholz column} \citep{G1990, L1987} and allowing 3D perturbations to the non-deformed filament. As a result, bifurcations to both planar flapping states and non-planar spinning motions have been found, reminiscent of the cone-like beating motion of cilia \citep{LGK2018}.

The ability of eukaryotic cilia to oscillate with distinct frequencies and wavelengths to control and manipulate fluids at the microscale has motivated the development of \textit{bio-inspired} synthetic macro-scale fluidic elements such as pumps or mixers consisting of soft filaments immersed in a fluid medium \citep{DR2019} to generate functionalities such as locomotion, mixing and mechanosensing due to elasto-hydrodynamic instabilities \citep{SG2020,WLT2022}. These active \textit{metasurfaces} of artificial cilia combine viscous dissipation and elasticity with generally \textit{non-conservative forces} originating from magnetic, electric or chemical fields to induce deformation, buckling and motion, thereby producing arbitrary flow patterns in fluids \citep{SG2020,WLT2022}. The works on bio-inspired flexible structures stimulated studies of 3D elasto-hydrodynamic instabilities such as flapping, swirling and flipping in prestressed, initially twisted filaments subject to active distributed follower forces and fluid drag \citep{FGG2018,FGG2021}.

\subsection{Spontaneous motion in growing  plants}

\begin{figure}[!ht]
\centering\includegraphics[width=\textwidth]{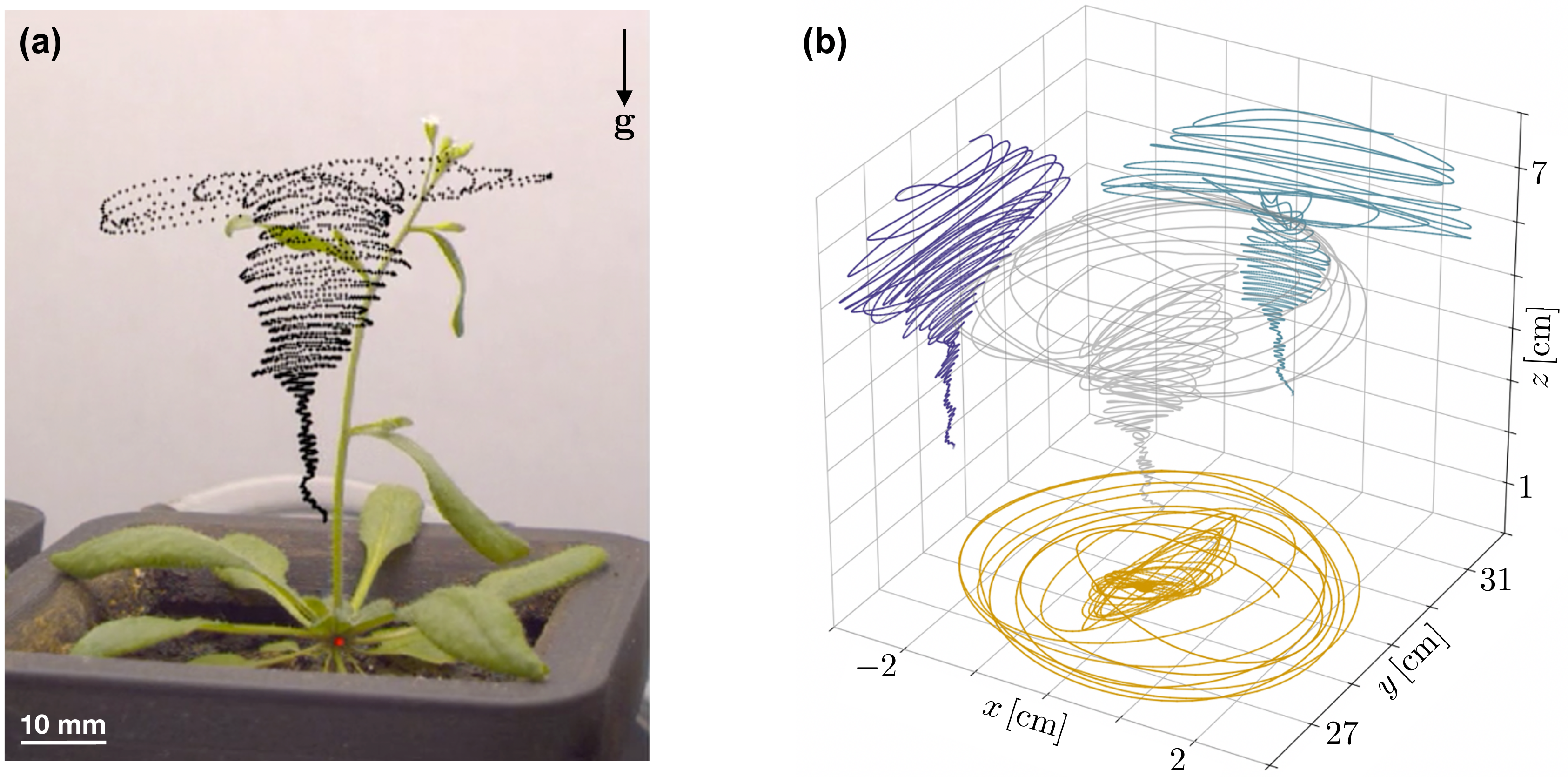}
\caption{Circumnutations in growing plant shoots. (a) Photo of a sample of \textit{Arabidopsis thaliana} (ecotype Col-0) exhibiting periodic oscillations under gravity, $\mathbf{g}$,  with superimposed plant tip trajectory. (b) Three-dimensional reconstruction of the experimental motion of the plant's tip. Notice the transition from small-amplitude, erratic movements to large-amplitude oscillations as the plant's shoot attains by growth the critical length associated with the onset of flutter instability.}
\label{fig_gels0}
\end{figure}

Another intriguing example of biological structures that exhibit spontaneous oscillations comes from growing plant shoots. These are slender active structures that, well before growing tall enough to become prone to buckling under their own weight \citep{G1881}, exhibit circular, elliptical, or pendular oscillations referred to as \textit{circumnutations} by Charles Darwin \citep{darwin_1880}. These movements are too slow to be detected by a casual observer, but they become apparent through time-lapse photography, Fig. \ref{fig_gels0}.
Plants are known to respond to environmental stimuli (as for instance light, gravity, and moisture) by means of tropic movements, responses of the plant organ that are typically driven by nonuniform tissue growth \citep{bastien_2014, chauvet_2019, meroz_2019, moulton_2022}. 
According to a prevailing theory for the interpretation of circumnutations, such oscillatory movements are of exogenous nature and arise from the over-compensatory response of the plant to the changing orientation of its gravi-sensory apparatus relative to the Earth’s gravity \citep{gradmann_1922, johnsson_1967, johnsson_1973, johnsson_2009, somolinos_1978}. Recently, mechanical models based on morphoelastic rod's theory \citep{G2017} have been proposed and tested to support this hypothesis \citep{AND2021b,ADN2021a,AL2020}.
Specifically, it was shown that nutations in growing plant shoots may be interpreted as spontaneous oscillations caused by a flutter instability (a Hopf bifurcation) in a growing elastic system subject to gravity loading, while sensing and actively responding to external stimuli. While endogenous factors contribute to the movements of growing plants, it was shown that their weight relative to those associated with flutter instability varies in time as the shoot length increases.
Also, within a linearized setting, the existence of oscillatory and diverging solutions above a critical length of the rod has been proven, thus extending a classical result by Greenhill \citep{G1881}. For a choice of material parameters consistent with the available literature on plant shoots, it was numerically found that, for active rods of sufficient length, pendular movements are unstable and eventually converge towards stable circular limit cycles, via transient elliptic trajectories of increasing amplitude. 
When compared with flutter in mechanical systems loaded by non-conservative (follower) forces \citep{BN2011}, the energy needed to sustain oscillations in the context of nutation of growing plant shoots is continuously supplied to the system by the internal biochemical machinery presiding the capability of plants to maintain a vertical pose.
%
%

\subsection{Mechanobiology and morphogenesis}

Mechanobiology and morphogenesis in living matter is another example of fundamental processes that involve mechanical instabilities \citep{FDC,G2017,XU,YIN}. 

With particular reference to flutter instability, an example is given by the periodic process in space and time of vertebral segmentation in an embryo. It is generally assumed that periodicity is controlled by a molecular `clock' which regulates temporally synchronized gene expression. Periodicity in space then emerges as a projection of the temporal periodicity performed by a steadily moving wave.  It was shown in \citep{TVS2014} that the primary periodicity may be of \textit{spatial} rather than temporal origin and that the required synchronization may be due to \textit{mechanical}, rather than biochemical, signalling. Mechanical instabilities can lead to the development of stress inhomogeneities in spatially distant material points thus generating a robust number of segments without
dependence on genetic oscillations \citep{ABC2017,TVS2014}.

\subsection{Human spinal column}

Leipholz \citep{L1987,LP1984} established the surprising result that a vertical cantilevered column under both its own weight and a compressive distributed follower load reaches a buckling load considerably higher than that corresponding to the same column in the absence of the follower load. He suggested verifying these results by means of laboratory experiments, noticing however that \lq because of difficulties in realizing steady follower forces, there has been little progress in this direction'. Leipholz's result finds an application to the mechanics of the ligamentous spinal column in the human body. In fact, the spine has long been known to support substantial compressive load \textit{in-vivo}, while the mathematical models and experiments show that, when  
subject to only axial dead load, the spine  
buckles at values of load almost an order of magnitude smaller than those typically loading the human lumbar spine during daily living activities   \citep{R2011,P2000, R2009}. Recent numerical works demonstrated that the distributed follower load can indeed be created in the human spine by spinal musculature \citep{R2011,R2009,Sun2022}, thus supporting the \textit{in-vitro} biomechanical tests results \citep{AFF2022,P2000} and the cited Leipholz's surprising result.

\section{Recent progress on flutter}

\subsection{The quest for follower forces}

The beginning and development of structural flutter were based on the invention of the concepts of follower torque and follower force, nonconservative actions applied at the end of a rod (rigid or elastic), remaining tangent to it during the deformation, as defined and used by Greenhill \citep{G1883,G1881}, Nicolai \citep{nicolai1929,nicolai1928}, Pfl\"uger \cite{Pfluger1955,Pfluger1950}, Beck \citep{B1952}, Ziegler \citep{Z1952,Z1953,Z1956,Z1977}, Bolotin \citep{B1963}, Herrmann and Jong \citep{herrmannjong}, Como \citep{como} and Leipholz \citep{L1987}. 
The presence of a follower force, even applied to simple elastic structures, shows many unexpected and counter-intuitive effects: (i.) lack of Eulerian instability; (ii.) presence of a Hopf bifurcation; (iii.) destabilizing effect of dissipation; (iv.) the so-called \lq Ziegler-Bottema paradox' \citep{HG2003,K2021,KV2022,KV2010,LFD2016,Z1952,Bottema1956}. 
These effects are so strange that the possibility of achieving a follower force, in reality, was severely criticized by Koiter and others, as explained in detail by Elishakoff \citep{E2005,E2020}. The basic idea beyond the criticism of the follower force was that the \lq strange effects' could be set aside and ignored in mechanics, would the follower force be a mathematical abstraction not reproducible in reality, using more explicit words, a fake model. 

\begin{figure}[!ht]
\centering\includegraphics[width=\textwidth]{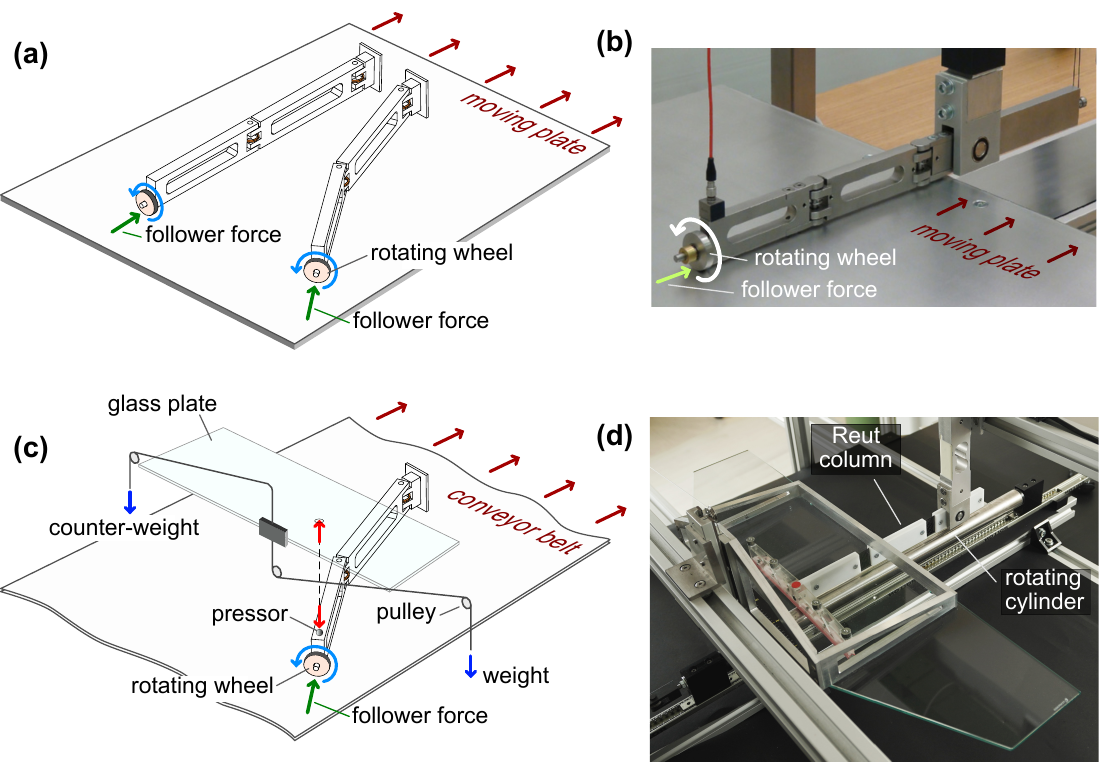}
\caption{
The realization of non-conservative forces: 
(a) and (b) follower forces obtained via Coulomb friction, acting on a freely rotating wheel sliding against a plate; (c) or sliding against a conveyor belt; (d) The Reut load obtained through the use of a freely rotating cylinder acting with friction on a double pendulum with a T-shaped end.
}
\label{Fig_01_flutter_machine}
\end{figure}

Several attempts to provide experimental evidence to the concept using air or fluid flow from a nozzle (Herrmann et al. \citep{Herrmann1966}, Saw and Wood \citep{SW1975}, Misseroni et al. \citep{misseroni2021extreme}) or using a rocket motor (Sugiyama et al. \citep{Sbook2019}, Uzny et al. \citep{UZNY2023}) were at least in part successful, nevertheless criticized from different points of view. 
A completely new approach was initiated by Bigoni and Noselli \citep{BN2011}, in which the follower force was created by exploiting Coulomb friction on a freely rotating wheel, as sketched in Fig.~\ref{Fig_01_flutter_machine}(a-b). The same device was later perfected \citep{BMTKN2018,BKMNT2018} 
(Fig.~\ref{Fig_01_flutter_machine}(c)) and the concept was also extended in \citep{BM2020} to reproduce another non-conservative (and strongly criticized) force, which acts on a fixed line, as postulated by Reut \citep{R1939}, Fig.~\ref{Fig_01_flutter_machine}(d). After more than fifty years of criticism and misunderstanding, the experiments performed at the {\it Instabilities Lab} of the University of Trento showed that follower forces are a correct mechanical model, reproducible using frictional contact interactions and that the Hopf bifurcation, dissipative instabilities, and Ziegler-Bottema paradox are all mechanical features measurable in a laboratory \citep{BKMNT2018,BMTKN2018}. 
It may be interesting to note that the way of obtaining the Ziegler's follower load introduced in \citep{BN2011} 
is in a sense \lq dual' to the technique designed to obtain the Reut's load \citep{BM2020}. In fact, in the former case a freely-rotating wheel attached to the structure is subject to a frictional force transmitted by a plate, while in the latter case, the plate is replaced by a freely-rotating tube, transmitting a frictional force to a part of the structure, which plays the role of the wheel in the former case.

The experimental set-up based on a moving plate \citep{BN2011} or a conveyor belt \citep{BMTKN2018,BKMNT2018} is appropriate to highlight another feature typical of flutter instability, leading through a Hopf bifurcation to a limit-cycle behaviour. Seen as a machinery powered through a steady source (the conveyor belt moving at constant speed), the structure reaches, after a remarkably short transitory,  a self-oscillatory state, in which the motion repeats itself with a fixed frequency, determined by the geometry, stiffness, damping characteristics of the system and by the velocity of the belt. 
Therefore, the \lq flutter machine' designed in \citep{BMTKN2018,BKMNT2018} represents an example of {\it self-oscillating device} \citep{jenkins2013self}.

\subsection{Oscillatory instability of piecewise smooth structures}
\label{apezzi}

\begin{figure}[!ht]
\centering\includegraphics[width=\textwidth]{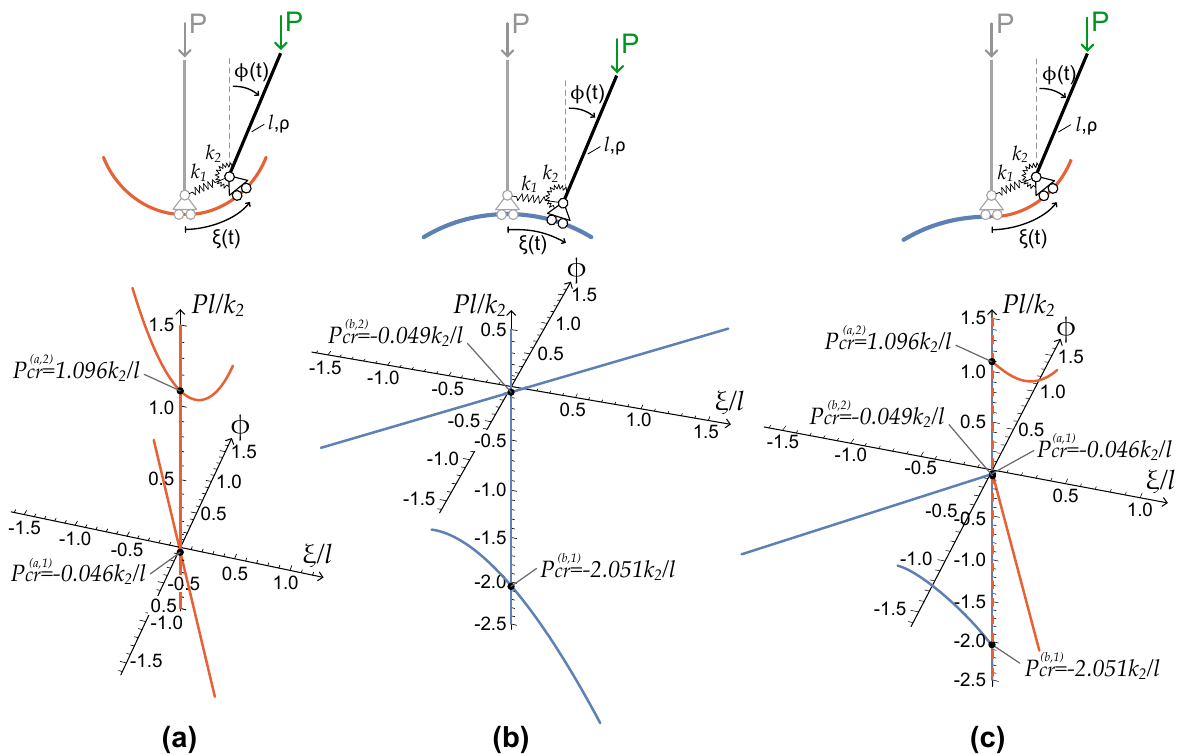}
\caption{
Two 2 d.o.f. structures loaded with a dead load (a) and (b) are \lq fused' together to obtain the compound structure (c). 
The rigid bar is connected with a rotational spring to the roller and with a longitudinal spring joining the roller to the undeformed position. 
The structures (a) and (b) have two bifurcation loads: one tensile and one compressive in the former case, both compressive in the latter. Structure (c), still with 2 d.o.f., exhibits 4 bifurcation loads, the set of all bifurcation loads pertaining to the two structures (a) and (b).
}
\label{fusione_1}
\end{figure}

Consider two elastic structures, each one composed of a rigid rod (of length $l$) elastically constrained (through a translational and a rotational spring of stiffnesses $k_1$ and $k_2$) to slide along a frictionless profile, smooth and with continuous curvature and {\it subject to a dead loading}, Fig. \ref{fusione_1}(a-b) upper part. The two structures differ in the curvatures of the sliding profile, having opposite signs, respectively $R_+>0$ and $R_-<0$. 
The lower parts of the figure show (for $R_+/l = 0.5$, $R_-/l = -1$, $k_1 l^2/k_2 = 0.1$) that, under conservative loading application, the structures have two bifurcation loads, respectively equal to $\{+1.096, \, -0.046\}\,k_2/l$ (where the \lq $+$' sign denotes a tensile buckling) and to $\{-0.049, \, -2.051\}\,k_2/l$, and include the post-critical behaviours.
Now imagine to \lq cut' the structures into two complementary halves and create a new structure joining the two different half-structures, as indicated in Fig. \ref{fusione_1}(c). 
%
%
The new structure still has 2 d.o.f. but is constrained by a frictionless profile with discontinuous curvature, leading to a discontinuous Hessian of the total potential energy. 
This discontinuity leads to multiple bifurcation loads for the compound structure shown in Figure (c), where a tensile and three compressive bifurcation loads exist, namely, exactly those arising in the continuous curvature structures (a) and (b) from which the discontinuous structure originates \citep{zak1, zak2}. 

\begin{figure}[!ht]
\centering\includegraphics[width=\textwidth]{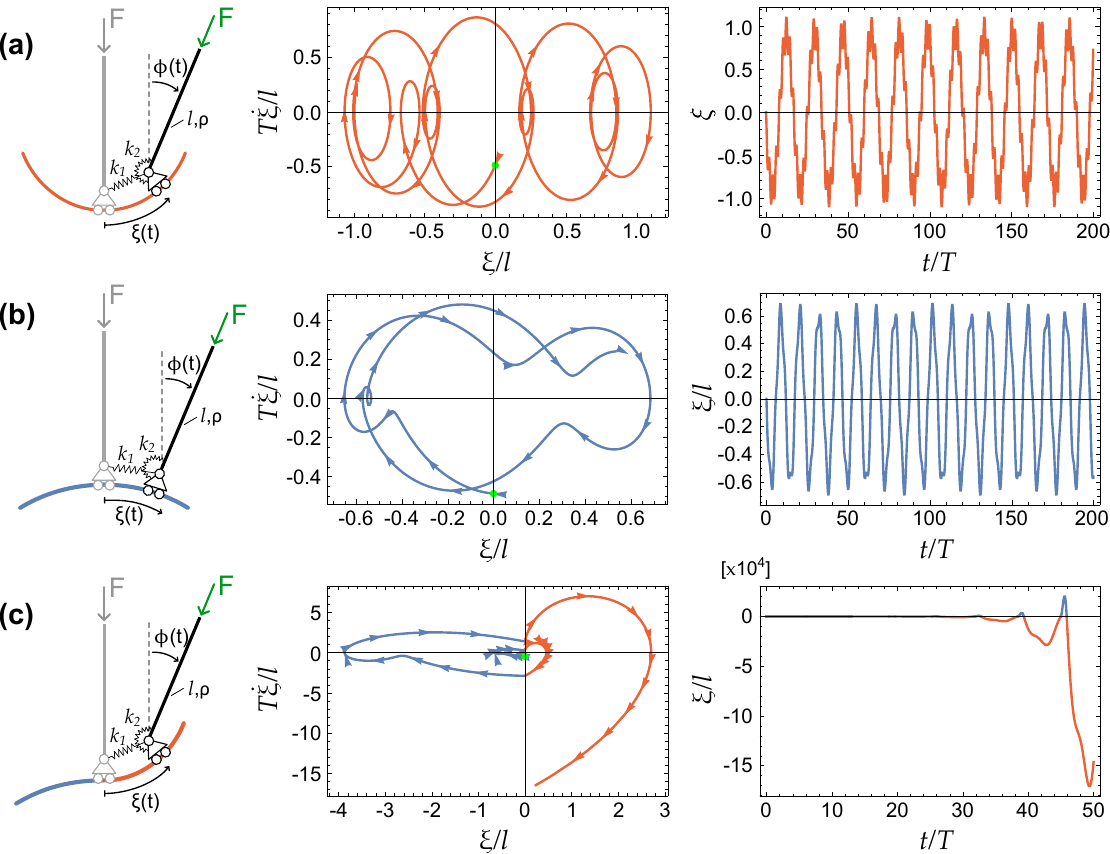}
\caption{
Two structures loaded within their stable range (a) and (b) are \lq fused' together to obtain the compound structure (c). Here the load is non-conservative and follower. The structure (c) is proven to be unstable at the same load for which the two generating structures are well within the stability region. Parameters used in this example are: $R_+ = 0.5\, l$, $R_- =- 2 R_+$, $k_1 = 0.1\, k_2/l^2$, $\rho = T^2 k_2/l^3$, $F = -2.4\, k_2/l$,
and $T$ is a reference time. 
}
\label{rossi}
\end{figure}

The described structural behaviour is apparently simple and related to the presence of loads following from a potential so that a much more complicated mechanical behaviour emerges when the {\it loads are follower}, as in the case shown in Fig. \ref{rossi}, where the applied force remains tangent to the rigid bar. 
%
%
In this case, quasi-static bifurcations do not occur, but the structures (a) and (b) are subject to flutter (respectively, $F_{flu}l/k_2=+1.10455$ and $F_{flu}l/k_2=-2.62091$) and divergence (respectively, $F_{div}l/k_2=+3.29545$ and $F_{div}l/k_2=-3.05909$) instabilities. The flutter instability is demonstrated in Fig.~\ref{rossi}, in terms of the phase diagram and with the behaviour in time shown for the Lagrangian parameter $\xi$, measuring the position of the roller along the sliding profile. 

The situation is more complex for the structure (c), for which the dynamics is governed by piecewise-smooth differential equations, of the type
\begin{equation}
\label{mutandine}
\begin{array}{ll}
    \boldsymbol{M}\ddot{\boldsymbol{q}} + 
    \boldsymbol{K}^- \boldsymbol{q} = 
    \boldsymbol{0}, ~~~
    \xi <0, \\
\boldsymbol{M}\ddot{\boldsymbol{q}} + 
    \boldsymbol{K}^+ \boldsymbol{q} = 
    \boldsymbol{0}, ~~~
    \xi >0,
    \end{array}
\end{equation}
where $\boldsymbol{M}$ is the mass matrix (equal for the two structures (a) and (b)), $\boldsymbol{K}^\mp$ are the two stiffness matrices (different for the two structures (a) and (b) and here both present), and $\boldsymbol{q}$ is the vector collecting the Lagrangean parameters $\xi$ and $\Phi$, both functions of time $t$. 

It may be anticipated that the flutter and divergence loads of the structures (a) and (b) remain critical for the compound structure (c), but this structure may be subject to unstable blowing-up oscillations, for a load smaller, $F l/k_2= -2.4$, than those leading to instability in the two separated structures (a) and (b) \citep{rossi}. This is shown in Fig. \ref{rossi}, where the oscillatory instability refers to the linearized system of equations and has been found with a technique based on the so-called \lq invariant cone' \citep{carmona, hosham, kupper, weiss1, weiss2}. The instability found for the linearized equations can be confirmed with fully nonlinear dynamical analyses. The instability is akin to the flutter instability, sharing with this the oscillatory blowing-up motion characterizing the unstable dynamics.

It is noted in the closure of this paragraph that 
the piece-wise linearity of the elastoplastic constitutive equations (\ref{platico}) makes the equations of the incremental dynamics of an elastoplastic solid similar to equations (\ref{mutandine}), governing the behaviour of the piece-wise smooth structure shown in Fig. \ref{rossi}.
Flutter in elastoplastic solids and in frictional contact between solids have usually been analyzed 
only within the framework of a linearized analysis, based on the plastic branch of the constitutive operator, equation (\ref{platico2}). The above-presented structural example shows that analyses based only on the plastic branch may fail to capture important instabilities \citep{petrykkiera}, while these seem to have been found in numerical simulations of dynamical plasticity of solid bodies \citep{brannon1, brannon2}. 
Therefore, the oscillatory instabilities found in the behaviour of the elastic structure shown in Fig. \ref{rossi} give evidence to possible deficiencies of stability investigations based on the concept of \lq comparison solid' 
\citep{bigonilibro}.

\subsection{Flutter in conservative systems}

Flutter and divergence instabilities, Hopf bifurcation, dissipation instabilities, and Ziegler-Bottema paradox are usually associated with the nonconservative nature of the follower load. 

This was the opinion, among others, of Anderson and Done \citep{anderson}, Huang et al. \citep{huangnc}, Singer et al.\ \citep{singer}. 
Elishakoff \citep{E2005} writes that \lq Bolotin felt [...] that it should be impossible to produce Beck's column experiment via a conservative system of forces'. 
Anderson and Done report that \lq conservative system can not become dynamically unstable since, by definition, it has no energy source from which to supply the extra kinetic energy involved in the instability'.

\begin{figure}[!ht]
\centering\includegraphics[width=\textwidth]{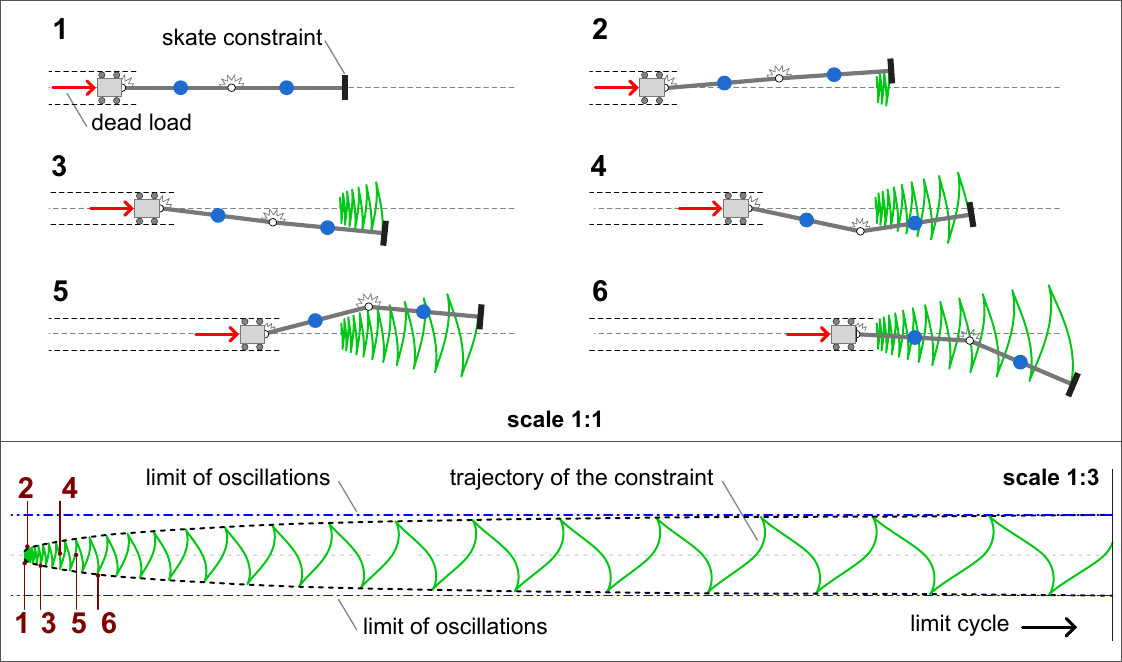}
\caption{Flutter instability, with Ziegler-Bottema paradox, in a 3 d.o.f. structure, subject to conservative (dead) loading and equipped with a nonholonomic skate constraint (on the left end). The sequence of deformations 1--6 shows the development of the Hopf bifurcation, with the attainment of a limit cycle (adapted from \citep{cazzolli1}).}
\label{Fig_02_flutter_holonomic}
\end{figure}

Disproving all the above claims and beliefs, a variant of the device invented by Bigoni and Noselli \citep{BN2011}, in which the freely rotating wheel becomes a non-holonomic constraint, impeding the velocity to be orthogonal to the wheel movement, shows that dynamic instabilities and effects related to the dissipation are possible in structures only subject to conservative forces \citep{cazzolli2,cazzolli1}. 
Fig.~\ref{Fig_02_flutter_holonomic} shows an example of a structure equipped with a non-holonomic constraint which, though different in the \textcolor{black}{nonlinear} dynamic response, has the same critical loads for flutter and divergence as the \lq Ziegler's double pendulum' and exhibit dissipation-induced instability and Ziegler-Bottema paradox.
\textcolor{black}{More specifically, the linearized equations of motion for this conservative system subject to a nonholonomic skate constraint  coincide with those for the \lq Ziegler's double pendulum', eqn \eqref{zieglinear}.}
Analogous examples have been provided for Reut's column, equipped now with a non-holonomic constraint \citep{cazzolli2,cazzolli1}.
It can be concluded that nonconservative forces are not necessary so that {\it a plethora of dynamic instabilities can simply be obtained using nonholonomic constraints, within a framework of conservative forces, easily achievable in a laboratory}.
%

Results on instabilities related to non-holonomic constraints were anticipated by works on the so-called \lq shimmy' and \lq caravan' instabilities, the former going back to 
Keldysh (1945) \citep{Keldysh} and the latter to Ziegler \citep{ SAF2002,TZ1984,Z1938}. 
Shimmy usually is an instability of a single wheel \citep{Stepan2, Pacejka, Plakhtienko, Smiley, Stepan1, Stepan3,vonSchlippe, Zhuravlev1, Zhuravlev2}, while caravan is about a 4-wheeled vehicle, coupled to a 2-wheeled trailer. The latter problem is more complicated than the former, because both the leading vehicle and the trailer oscillate in a way similar to the \lq Ziegler's double pendulum' \cite{Z1938,BEREGI2016}.

\subsection{Metamaterials and non-Hermitian elasticity}

In order to enforce the validity of the rules of thermodynamics, elasticity is normally based on a strain potential and  called \lq hyperelasticity', a concept dating back to 
George Green and Lord Kelvin \citep{green, kelvin2,kelvin1}. In linear elasticity, where the stress $\sigma_{ij}$ and the strain $\varepsilon_{ij}$ are related through a fourth-order tensor $\mathcal{E}_{ijhk}$ as
\begin{equation}
    \sigma_{ij} = \mathcal{E}_{ijhk} \varepsilon_{hk}, 
\end{equation}
a necessary and sufficient condition for the existence of strain potential $\Phi$ is the major symmetry
\begin{equation}
    \mathcal{E}_{ijhk} = \mathcal{E}_{hkij},
    ~~~
    \mbox{ so that }
    ~~~
    \Phi = \frac{1}{2} \varepsilon_{ij}{\mathcal{E}}_{ijhk}\varepsilon_{hk}  
    ~~~
    \mbox{ and }
    ~~~
    \sigma_{ij} = \frac{\partial \, \Phi}{
    \partial \varepsilon_{hk}}.
\end{equation}
An elasticity tensor lacking the major symmetry defines a \lq non-Hermitian material' or a \lq Cauchy-elastic' or \lq hypo-elastic material' without strain potential \citep{truesdell, truesdell-noll}. 
In this condition, the material can always produce or absorb energy when subject to an appropriate, closed, loop of strain. Obviously, energy production is impossible, so these materials borrow (or lend) the energy from (or to) some external source. 
Examples of \lq non-Hermitian systems' (rather than materials) have been introduced in the dynamics of metamaterials, to denote {\it active} materials (thus containing motors, actuators, or piezoelectric devices), which can exchange energy with their surroundings and in this way, {\it apparently} producing energy in a closed strain cycle 
\citep{attarzadeh, chen, Coulais2021, Ghatak2020, rosa, vitelli, torrent1, torrent2, wang, correntin}. 

\begin{figure}[!ht]
\centering\includegraphics[width=\textwidth]{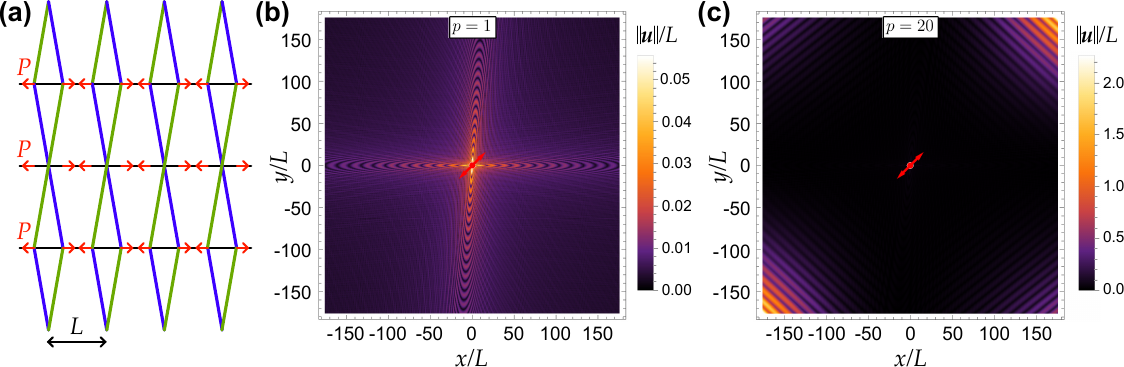}
\caption{
Incremental displacement maps for a non-Hermitian hypoleastic material excited with a pulsating point force, located at the centre of the domain and inclined at 45$^\circ$. (a) Grid of elastic rods loaded with follower microforces $P$ (shown red) to which the non-Hermitian material is equivalent.
The properties of the rods are as follows: $\gamma_j, A_j, B_j$ represent the linear mass density, axial stiffness, and bending stiffness of the rods, respectively. The rods are coloured black, green, and purple corresponding to $j=1,2,3$. The angle of inclination of the green and purple rods with respect to the horizontal axis is denoted by $\alpha$. (b) A stable behaviour, showcasing localized vibrations due to elastic anisotropy is obtained when $p = PL^2/(2B_1)=1$ (where $L$ is the length of the horizontal rods). (c) Flutter instability is obtained when $p = 20$. Flutter
results in unbounded wave growth in space, degenerating into planar waves of characteristic inclination (visible near the corners of the figure and inclined at 45$^\circ$). Parameters used: $\gamma_2 = 3 \gamma_1/10$, $\gamma_3 = \gamma_1/25$, $A_2 = A_1$, $A_3 = 25 A_1$, $B_1 = A_1 L^2 / 6400$, $B_2 = A_2 L^2 / (25600 \cos^2\alpha)$, $B_3 = A_3 L^2 / (78400 \cos^2\alpha)$, $\alpha = 80^\circ$.
}
\label{flutter_continuo}
\end{figure}

A way to obtain a non-Hermitian elastic material has been indicated by Bordiga et al.\ \citep{bordiga} through a rigorous homogenization of a periodic grid of elastic rods (deformable under axial and shear forces and bending moment) subject to {\it follower, periodically distributed, microforces}, as shown in Fig. \ref{flutter_continuo} (a). The energy exchange with the surroundings is provided by the microforces. 

Similarly to the definition (\ref{acuplat}), a non-Hermitian material leads to a non-symmetric (second-order) acoustic tensor
\begin{equation}
\label{acuelat}
    A_{ij}(\mathbf{n}) = \mathcal{E}_{ihjk}n_h n_k, 
\end{equation}
defined for the unit vector $\mathbf{n}$, ruling the wave propagation direction.  
Again similarly to nonassociative elastoplasticity, the acoustic tensor (\ref{acuelat}) can have two complex conjugate eigenvalues, defining flutter instability in a continuum. Indeed this instability was found in the homogenized response of the non-Hermitian elastic material obtained by Bordiga et al.\ \citep{bordiga} through the homogenization of a grid of elastic rods, Fig. \ref{flutter_continuo} (a).
%
%
In this way, a formal link has been discovered between structural flutter and its occurrence in a continuum. 
The figure demonstrates that flutter in a continuum corresponds to an unbounded growth of waves (emanating from a pulsating point force, represented as a double-headed arrow and located at the centre of the figure), degenerating into planar waves with a well-defined inclination. A stable behaviour is shown in panel (b), while flutter, occurring at a higher value of follower microforces, is shown in panel (c). 
Growth, instead of decay, of wave amplitude is a consequence of the fact that the wave is able to absorb energy from the environment during its propagation. 
Therefore, the realization of a non-Hermitian elastic material would lead to devices capable of transmitting mechanical waves not only without attenuation but rather with amplification, a finding which foreshadows the possible development of a mechanical laser.

\subsection{Flutter-based actuation for soft robotics}

\begin{figure}[!ht]
\centering\includegraphics[width=\textwidth]{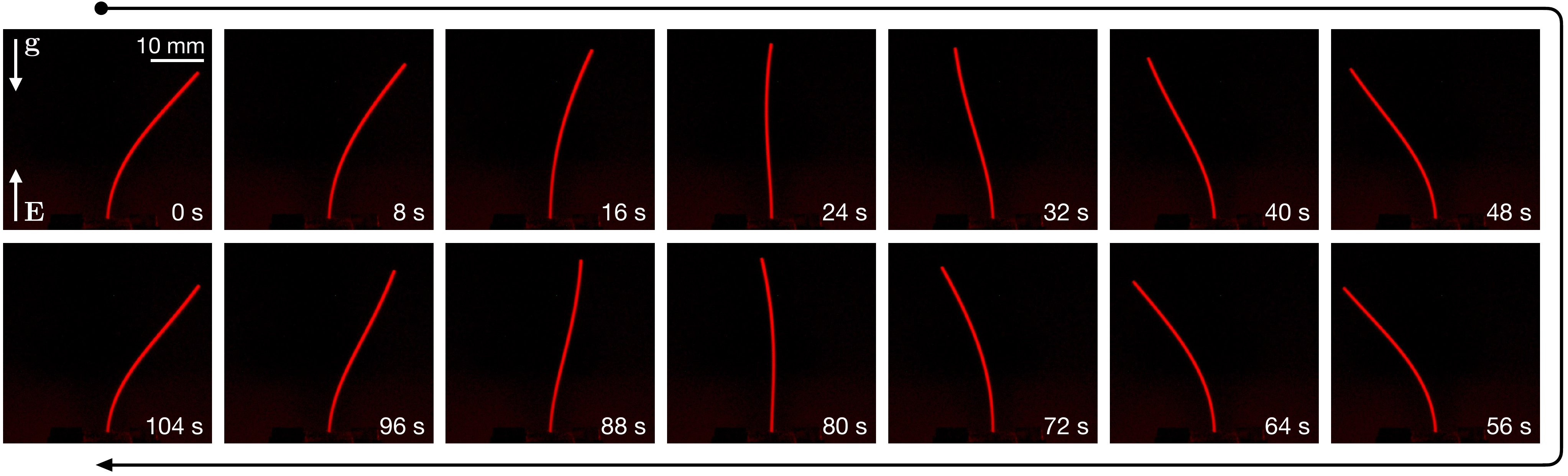}
\caption{A sequence of snapshots showing a polyelectrolyte hydrogel rod exhibiting periodic, self-sustained oscillations under the combined action of gravity, $\mathbf{g}$, and of a uniform and steady electric field, $\mathbf{E}$. The snapshots were taken 8\,s apart.}
\label{fig_gels}
\end{figure}

The active reconfigurations and movements of elongating plant organs have recently received renewed attention for the design and optimal actuation of bioinspired \textit{soft robotics} actuators \citep{A2023,M2022,M2023}.
In particular, as shown in Fig. \ref{fig_gels}, it was demonstrated that active filaments of polyelectrolyte gels operating in a viscous environment can exhibit periodic oscillations arising from flutter or divergence instability, when subject to a steady and uniform electric field, with application to smart actuator design for micro-fluidics or soft-robotics \citep{CDN2023}. In such a physical system, the externally applied electric field, which causes the spontaneous bending of the gel samples from differential swelling, plays a similar role to that of Earth’s gravitational field in plant’s tropism, where gravity causes the active reaction of plant shoots driven by differential growth. In this regard, previous studies \citep{stone_2021,stone_2020} have provided evidence that locomotion of artificial swimmers consisting of elastic filaments attached to a dielectric particle can be achieved under a uniform and static electric field by a Hopf's bifurcation driven by Quincke rotation.
%
%
In a related context, cyclic motions in films of liquid crystal elastomers have been shown to occur under steady illumination \citep{gelebart_2017,yamada_2008} and then interpreted within a unifying framework based on a nonlinear model for photo-deformable rods \citep{korner_2020}.

\section{Conclusion and outlook}

Research on flutter instability is an intriguing and fully vital subject, with applications in an extremely broad range of fields, including the dynamics of slender bodies, instability of machines and wheel systems, fracture propagation, fault slip, granular matter, non-Hermitian materials, metamaterials, biomechanics, growth, animal and robotic locomotion. 

On this subject, many aspects of fundamental scientific importance are still to be discovered and only recent progress has been made. 
Applying follower forces on architected materials and metamaterials poses significant challenges. Currently, the only viable methods are through dry friction (non-holonomic constraints are in a sense included as the case of infinite friction) or, with certain approximations, via fluid flow. However, the latter technique introduces complexities due to fluid-structure interactions, making difficult the precise control of the applied follower force. Moreover, miniaturizing friction-based devices for implementing follower forces at different locations within architected materials is highly complex. In order to significantly progress in this field, a dedicated research effort should be directed towards devising innovative approaches for applying follower forces within architected materials, metamaterials and structures. These approaches should not only provide enhanced precision and control over the applied forces, but also address the associated technological challenges of miniaturization. Moreover, the interplay between configurational forces, realized in time-varying length structures, and flutter mechanisms has never been considered. By exploring and developing novel techniques and structural schemes for the application of follower forces, the potential of architected materials and metamaterials can be fully unlocked, thereby opening up groundbreaking possibilities in various domains, such as energy harvesting.

Extensive research has been conducted in aerospace engineering and fluid dynamics to study the flutter phenomenon. However, its application in energy harvesting is a relatively new and promising research area, generating significant interest and advancements. The potential of flutter energy harvesters to exploit wind and fluid flows in innovative ways presents exciting opportunities for the renewable energy sector. To maximize energy harvesting through flutter instabilities, it is imperative to develop lightweight, flexible architected materials and advanced sensor and energy conversion technologies.
The progress in this field is crucially related to continued research, multidisciplinary collaboration, and technological advancement. Efforts in this direction will lead to improved understanding, optimized design, and practical realization of flutter-based energy harvesters. 

Additionally, combining experimental investigations and computational simulations will provide deeper insights into the physics of flutter instabilities and will possibly lead to the discovery of new technologies, based on non-Hermitian materials. These can be related to the conditioning and manipulation of mechanical signals, possibly leading to propagation with amplification or to the design of a mechanical laser.

Furthermore, it is essential to develop control strategies to mitigate the risk of destructive vibrations and ensure optimal performance for materials and structures subjected to oscillatory instabilities. 
In fact, flutter has always been associated with structural breakdown and the oscillations that led to the Tacoma bridge failure are evidence of this phenomenon. 
Flutter in continua also represents an important instability occurring in granular media, rocks, and fault slip.  Several phenomena are here still awaiting explanation, so the research in this field may have strong implications in the processing of granular matter, in the stability of slopes, and in earthquake prediction as related to unstable fault slips.

The long-term durability and reliability of materials and structures exposed to dynamical instabilities must be thoroughly studied, considering environmental factors such as fatigue, and extreme weather conditions, which can impact their structural integrity.

\section*{Acknowledgements}
D.B., F.D.C., D.M., A.P. acknowledge financial support from the European Research Council (ERC) under the European Union’s Horizon 2020 research and innovation programme (Grant agreement No. ERC-ADG-2021-101052956-BEYOND). G.N. acknowledges financial support from the Italian Ministry of University and Research (MUR) through the grant \lq Dipartimenti di Eccellenza 2018-2022 (Mathematics Area)'.

\printbibliography

\end{document}